\documentclass[prl,superscriptaddress,showpacs,amssymb,letterpaper,twocolumn]{revtex4}

\usepackage{amsmath}
\usepackage{amsfonts}
\usepackage{graphicx}

\def\>{\rangle}
\def\<{\langle}

\def\be{\begin{equation}}
\def\ee{\end{equation}}

\begin{document}
\title{Quantum protocol for cheat-sensitive weak coin flipping}

\author{R. W. Spekkens}
\email{spekkens@physics.utoronto.ca}
\affiliation{University of Toronto, 60 St.George Street, Toronto M5S
1A7,
Canada}

\author{Terry Rudolph}
\email{rudolpht@bell-labs.com}
\affiliation{Bell Labs, 600-700 Mountain Ave., Murray Hill, NJ 07974, U.S.A.}
\affiliation{Institut f\"ur Experimentalphysik, Universit\"at Wien, Boltzmanngasse 5,
1090 Vienna, Austria}


\begin{abstract}
We present a quantum protocol for the task of weak coin flipping.
We find that, for one choice of parameters in the protocol, the
maximum probability of a dishonest party winning the coin flip if
the other party is honest is $1/\sqrt{2}$. We also show that
if parties restrict themselves to strategies wherein they
cannot be caught cheating, their maximum probability of winning can
be even smaller.  As such, the protocol offers additional security in
the form of cheat sensitivity.
\end{abstract}~


\maketitle

In 1981 Blum\cite{blum} introduced the following cryptographic
problem: Alice and Bob have just divorced and are trying to
determine who will keep the car. They agree to decide the issue
by the flip of a coin, but they can only communicate by
telephone. The question is whether there is a protocol that
allows them to decide on a winner in such a way that both parties
feel secure that the other cannot fix the outcome.

Two-party protocols, of which this is an example, are some of the
most problematic in classical cryptography. In fact, there are no
two-party classical protocols whose security does not rely upon
assumptions (many of which are threatened by quantum computation)
about the complexity of a computational task. Kilian explains
\cite{kilian}:
\begin{quote}
{[In a two-party protocol] both parties possess the entire
transcript of the conversation that has taken place between them.
[\ldots] Because of this knowledge symmetry condition there are
impossibility proofs for seemingly trivial problems.
Cryptographic protocols ``cheat'' by setting up situations in
which A may determine exactly what B can infer about her data,
from an information theoretic point of view, but does \emph{not}
know what he can easily (i.e. in probabilistic polynomial time)
infer about her data. \emph{From an information theoretic point of
view, of course, nothing has been accomplished.}(emphasis added)}
\end{quote}
Conversely, when we move from classical to quantum cryptography,
we find  many two-party protocols whose security rests only upon
the validity of quantum mechanics.  Thus, from a quantum
information-theoretic point of view, something significant
\textit{can} be accomplished. Furthermore, quantum protocols can
naturally exhibit a type of security known as
\textit{cheat sensitivity} \cite{HardyKent}: whenever a party
cheats above some threshold amount, he or she runs a risk of
being caught. This can provide a strong deterrent to cheating.
For instance, if two parties need to implement a protocol many
times, they may stand to gain more from the preservation of the
trust of the other party than they do from cheating in a single
implementation. Such considerations can be treated quantitatively
by assigning numerical costs to the various possible results.  Given
the striking contrasts between what can be accomplished in
classical and quantum two-party protocols, the analysis of such
protocols provides valuable insights into the differences between
classical and quantum information theory.

In this letter, we will be concerned with a cryptographic task
called \emph{coin flipping}. We begin by distinguishing a strong
and a weak form, both of which are adequate for Blum's original
problem.

\noindent\textit{Strong Coin Flipping}(SCF): Alice and Bob engage in
some number of rounds of communication, at the end of which each
infers the outcome of the protocol to be either $0$, $1$, or
\textit{fail}. If both are honest then they agree on the outcome and
find it to be $0$ or $1$ with equal probability. Suppose, on the
other hand, that one of the parties, $X$, is dishonest. In this
situation, $X$ cannot increase the probability of his/her
opponent obtaining the outcome $c$ to greater than $1/2+\epsilon
_{X}^{c}$, for either $c=0$ or $c=1$.
The parameters $\epsilon_{A}^{0},\epsilon _{A}^{1},\epsilon
_{B}^{0},\epsilon _{B}^{1},$ which specify the degree to which the protocol resists biasing, must each be strictly less than $1/2.$

\noindent\textit{Weak Coin Flipping}(WCF): This is simply SCF without any constraints
on $\epsilon _{A}^{0}$ or $\epsilon _{B}^{1}$. The parameters
$\epsilon _{A}\equiv \epsilon _{A}^{1}$ and $\epsilon _{B}\equiv
\epsilon _{B}^{0}$ must be strictly less than 1/2 and specify the bias-resistance of the protocol.

An SCF protocol ensures that neither party can fix the outcome to
be $0$ or fix the outcome to be $1.$ This protocol is appropriate
when the parties do not know which outcome their opponent favors.
By contrast, a WCF protocol only ensures that Alice
cannot fix the outcome to be $1$ and that Bob cannot fix the
outcome to be $0.$ This is appropriate if Alice and Bob are
playing a game where Alice wins if the outcome is $1$ and Bob
wins if the outcome is $0.$

It has been shown by Lo and Chau \cite{lochau} that a
\textit{perfectly bias-resistant} SCF protocol, \textit{i.e.} one having
$\epsilon _{A,B}^{0,1}=0$, is impossible. Recently, Kitaev
\cite{kitaev} has shown that it is
also impossible to find an \textit{arbitrarily bias-resistant} SCF protocol,
\emph{i.e.} one
for which $\epsilon _{A,B}^{0,1}\rightarrow 0$ in the limit that some
security parameters go to infinity. The first
\textit{partially bias-resistant} SCF protocol,
presented by Aharonov
\textit{et al.}\cite {aharonov}, had $\epsilon_{B}^{0,1}\simeq 0.354$ and $\epsilon_{A}^{0,1}\le 0.414$. We later showed that $\epsilon_{A}^{0,1}=\epsilon_{B}^{0,1}$ \cite{qic}.
If $\epsilon _{A}^{c}=\epsilon _{B}^{c}$ for $c=0$ and $1$, we call the protocol
\textit{fair}; if $\epsilon _{X}^{0}=\epsilon _{X}^{1}$ for $X=A$ and $B$, we call it \textit{balanced}.
A fair and balanced SCF protocol
with $\epsilon_{A,B}^{0,1}=\frac{1}{4}$ was recently discovered by Ambainis \cite{ambainis}; the possibility of SCF with this degree of security also follows from our analysis \cite
{spekkens} of quantum bit commitment. In fact, the results of
Ref. \cite {spekkens} imply the existence of a balanced SCF protocol
with $\epsilon _{A}^{0,1}=\alpha $ and $\epsilon
_{B}^{0,1}=\beta $ for any pair of values $\alpha,\beta $ satisfying $\alpha +\beta =1/2.$

Much less is known about WCF.
Indeed, whether arbitrarily bias-resistant WCF is possible or not remains an
open question.
Since an SCF protocol yields a WCF protocol with parameters
$\epsilon_{A}=\epsilon_{A}^{0}$ and $\epsilon_{B}=\epsilon_{B}^{1}$,
the protocol of Ref. \cite{spekkens} yields a WCF protocol with
$\epsilon_{A}+\epsilon_{B}=1/2$. However, it is likely that by
making a SCF protocol unbalanced one can lower the values of
$\epsilon_{A}^{1}$ and $\epsilon_{B}^{0}$ at the expense of
$\epsilon_{A}^{0}$ and $\epsilon_{B}^{1}$.  Thus, one would
expect there to exist a WCF protocol with better security than
the one derived from Ref. \cite{spekkens}. This expectation is
borne out by the results of this letter.  Specifically, we
demonstrate the existence of a three-round WCF protocol for
any $\epsilon_A$, $\epsilon_B$ satisfying
$(1/2+\epsilon_A)(1/2+\epsilon_B)=1/2$. In particular, this
implies that there exists a fair WCF
protocol with $\epsilon _{A,B}=(\sqrt{2}-1)/2\simeq 0.207$.

We also characterize the cheat sensitivity of this protocol.
Specifically, we consider each party's \textit{threshold for cheat sensitivity}, defined as the maximum probability of
winning that the party can achieve while ensuring that his or her
probability of being caught cheating remains strictly zero. Since
a party can achieve a probability of winning of 1/2 without
cheating, the minimum possible threshold is 1/2. The maximum
possible threshold is simply the party's maximum probability of
winning. The protocol is only said to be cheat-sensitive if the
threshold is less than this maximum value. We find that for
suitably chosen parameters, the protocol presented here can be cheat-sensitive against both parties simultaneously.  Although no parameter choices yield a threshold of 1/2 for both parties simultaneously, it is possible to obtain such a threshold for one of the parties.

\noindent\textbf{The protocol:}

\noindent\textit{Round 1.} Alice prepares a pair of systems in a (typically
entangled) state $\left|\psi\right\rangle\in\mathcal{H}^{A}\otimes \mathcal{H%
}^{B},$ and sends system $B$ to Bob.

\noindent \textit{Round 2.} Bob performs the measurement associated with the
positive operator-valued measure (POVM) $\left\{ E_{0},E_{1}\right\} $ on
system $B,$ and sends a classical bit $b$ indicating the result to Alice.

\noindent \textit{Round 3.} If $b=0$ then Bob sends system $B$ back to
Alice, while if $b=1$ then Alice sends system $A$ to Bob. The
party that receives the system then performs the measurement
associated with the projection valued measure $\{|\psi
_{b}\>\<\psi _{b}|,I-|\psi _{b}\>\<\psi _{b}|\}$, where $\left|
\psi _{b}\right\rangle =I\otimes \sqrt{E_{b}}\left|
\psi \right\rangle /\sqrt{\left\langle \psi \right| I\otimes E_{b}\left|
\psi \right\rangle }$. The different possible outcomes are:

\noindent {(i)} $b=0,$ Alice finds $\left| \psi _{0}\right\rangle\<\psi_{0}| $; Bob
wins.

\noindent {(ii)} $b=0$, Alice finds $I-\left| \psi _{0}\right\rangle\<\psi_{0}| $; Alice
catches Bob cheating.

\noindent {(iii)} $b=1,$ Bob finds $\left| \psi _{1}\right\rangle\<\psi_{1}| $; Alice
wins.

\noindent {(iv)} $b=1,$ Bob finds $I-\left| \psi
_{1}\right\rangle\<\psi_{1}|$; Bob catches Alice cheating.

Notice that unlike other proposed two-party protocols, at no
stage does this protocol require either party to make classical
random choices. While this protocol is sufficient for WCF, it is
insufficient for SCF because Bob can always choose to lose by
simply announcing $b=1$. We will see that one can characterize an
instance of the protocol completely by specifying the POVM
element $E_{0}$ and the reduced density operator on system $B$, $\rho \equiv
\mathrm{Tr}_{A}(\left|\psi \right\rangle\left\langle \psi \right| )$. In order for the
parties to have equal probabilities of winning when both are
honest, the constraint $\mathrm{Tr}\left( \rho
E_{0}\right) =1/2$ must be satisfied. This implies, in particular,
that $\left| \psi _{b}\right\rangle =\sqrt{2}\left( I\otimes \sqrt{E_{b}}%
\left| \psi \right\rangle \right).$

We proceed by listing the most important properties of the protocol. We then
present several interesting specific choices of $E_{0}$ and $\rho $. The
proofs are left until the end.

\noindent \emph{\textbf{Property 1:}} Alice's maximum probability of winning is
\[
P_{A}^{\max }=2\mathrm{Tr}\left( \rho E_{0}^{2}\right) \newline
\]

\noindent \emph{\textbf{Property 2:}} Alice's threshold for cheat
sensitivity is
\[
P_{A}^{\text{thresh}}=\frac{1}{2\mathrm{Tr}\left(
\rho \Pi _{(I-E_{0})}\right) },
\]
where $\Pi _{X}$ denotes the projector onto the support of $X$ (the support
of $X$ is the set of eigenvectors of $X$ associated with non-zero
eigenvalues).

\noindent \emph{\textbf{Property 3:}} Bob's maximum probability of winning
is
\[
P_{B}^{\max }=2(\mathrm{Tr}\sqrt{\rho E_0 \rho}) ^{2},
\]

\noindent \emph{\textbf{Property 4:}} Bob's threshold for cheat sensitivity
is
\[
P_{B}^{\text{thresh}}=\frac{1}{2\lambda ^{\max }\left( E_{0}\Pi _{\rho
}\right) },
\]
where $\lambda ^{\max }(X)$ denotes the largest eigenvalue of $X$.

An interesting family of protocols is defined by the choices
$\rho =x|0\>\<0|+(1-x)|1\>\<1|$ and $E_{0}=\frac{1}{2x}|0\>\<0|$,
where $1/2<~x~\le 1$.
For these protocols, $P_{A}^{\max }=1/2x$, $P_{B}^{\max }=x$, $P_{A}^{\text{%
thresh}}=1/2$, $P_{B}^{\text{thresh}}=P_{B}^{\max }$. Thus Alice
runs a risk of being caught whenever she cheats, while Bob can
cheat up to the maximum amount possible without running any
risk of being caught. This family achieves the trade-off
\begin{equation}
P_{A}^{\max }P_{B}^{\max }=1/2.  \label{tradeoff}
\end{equation}

It is easy to prove that this trade-off is optimal when $E_0$ and
$\rho$ have support in a 2-d Hilbert space.  In a preprint version
of this letter, we conjectured that it was optimal for all higher
dimensional Hilbert spaces as well. Subsequently, this was proven by
Ambainis \cite{ambainislowerbound} (who also independently discovered
a WCF protocol achieving the trade-off of Eq.(\ref{tradeoff})).
It is interesting to note that whereas
the best known SCF protocols \cite{ambainis, spekkens} require a
qutrit for their implementation, a qubit suffices here.

A second interesting family of protocols is defined by the choices
$\rho =x|0\>\<0|+(1-x)|1\>\<1|$ and
$E_{0}=(1-\frac{1}{2x})|0\>\<0|+|1\>\<1|$, with $1/2\le x<1$. For
these, $P_{A}^{\max }=1/2x$, $P_{B}^{\max
}=2+4x^{2}-5x+2(1-x)\sqrt{2x(2x-1)}$,
$P_{A}^{\text{thresh}}=P_{A}^{\max } $,
$P_{B}^{\text{thresh}}=1/2$. In contrast with the previous
example, Bob now runs a risk of being caught whenever he cheats,
while Alice can cheat up to the maximum amount possible
without running any risk of being caught. The trade-off
(\ref{tradeoff}) is no longer attained however.

It can be shown that no
choice of $E_{0}$ and $\rho$  can give
$P_{A}^{\text{thresh}}=P_{B}^{\text{thresh}}=1/2$ \cite{2wayCS}.
Nonetheless, it \textit{is} possible to have
$P_{A}^{\text{thresh}}<P_{A}^{\max }$ and
$P_{B}^{\text{thresh}}<P_{B}^{\max }$, \textit{i.e.}, cheat sensitivity against both parties simultaneously. This occurs, for example,
when $\rho=\frac{1}{2}I$ and
$E_{0}=\frac{3}{4}|0\>\<0|+\frac{1}{4}|1\>\<1|$, since in this
case $P_{A}^{\max }=5/8$, $P_{A}^{\text{thresh}}=1/2$,
$P_{B}^{\max}=\frac{1}{2}+\frac{\sqrt{3}}{4}\simeq 0.933$, and
$P_{B}^{\text{thresh}}=2/3$. In this case, if the parties restrict themselves to strategies wherein they cannot be caught cheating, their maximum probability of winning is even less than $1/\sqrt{2}$.  This example demonstrates that cheat sensitivity is a useful form of security in its own right.


\noindent\emph{\textbf{Proof of Property 1:}} Assume that Bob is honest.
Alice's most general cheating strategy is to prepare a state $|\psi ^{\prime
}\>$ instead of the honest $\left| \psi \right\rangle .$ (It is obvious from
what follows that she gains no advantage by preparing a mixed state, and
thus no advantage by implementing strategies wherein she performs
measurements on $A$ or entangles $A$ with a system she keeps in her
possession. Moreover, since she only submits $A$ to Bob when $b=1,$ any
operation on $A$ she wishes to perform can be done prior to Bob's
announcement, and thus can be incorporated into the preparation.)
The probability that Bob obtains the outcome $b=1$ is $\left\langle \psi
^{\prime }\right| I\otimes E_{1}\left| \psi ^{\prime }\right\rangle ,$ and
the probability that Alice passes Bob's test for $\left| \psi
_{1}\right\rangle $ when she resubmits system $A$ is $\left| \left\langle
\psi _{1}|\psi _{1}^{\prime }\right\rangle \right| ^{2},$ where $|\psi
_{b}^{\prime }\>\equiv (I\otimes \sqrt{E_{b}}|\psi^{\prime }\>
)/\sqrt{\<\psi^{\prime } |I\otimes E_{b}|\psi^{\prime }\> }$. Alice only
wins the coin flip if the outcome is $b=1$ \emph{and }she passes
Bob's test. This occurs with probability $P_{A}=\left\langle \psi
^{\prime }\right| I\otimes E_{1}\left|
\psi ^{\prime }\right\rangle \left| \left\langle \psi _{1}|\psi _{1}^{\prime
}\right\rangle \right| ^{2}$ $=|\<\psi _{1}|I\otimes
\sqrt{E_{1}}|\psi ^{\prime }\>|^{2}.$ We wish to find
$P_{A}^{\max }\equiv \sup_{|\psi ^{\prime }\>}P_{A}$. Thus, we
must maximize the overlap of a normalized
vector $|\psi ^{\prime }\>$, with the non-normalized vector $I\otimes \sqrt{%
E_{1}}|\psi _{1}\>$. Clearly, this is done by taking the two vectors
parallel, so the optimal $\left| \psi ^{\prime }\right\rangle $ is $|\psi
^{\prime }{}^{\max }\>=\left( I\otimes \sqrt{E_{1}}|\psi _{1}\>\right) /%
\sqrt{\<\psi _{1}|I\otimes E_{1}|\psi _{1}\>}$. Using the definition of $\left|
\psi _{1}\right\rangle $ and applying some straightforward
algebra, we find $P_{A}^{\max }=2\mathrm{Tr}(\rho E_{1}^{2})$. As $%
E_{1}^{2}=(I-E_{0})^{2}$ we obtain $P_{A}^{\max
}=2\mathrm{Tr}(\rho E_{0}^{2})$.\hfill $\blacksquare $

\noindent\emph{\textbf{Proof of Property 2:}} We seek to determine Alice's
maximum probability of winning assuming that her probability of
being caught cheating is strictly zero. Alice's most general
cheating strategy is, as above, to prepare a pure state $\left|
\psi ^{\prime }\right\rangle .$ She must pass Bob's test with
probability one, which implies $\left| \left\langle \psi _{1}|\psi
_{1}^{\prime }\right\rangle \right| ^{2}=1,$ or $\left| \psi
_{1}^{\prime }\right\rangle =\left| \psi _{1}\right\rangle $ to
within a phase factor. Multiplying both sides of this latter equation by $I\otimes \sqrt{%
E_{1}}^{-1}$ (we use $X^{-1}$ to denote the inverse of $X$ on its support), and writing $\left| \psi _{1}^{\prime }\right\rangle $ and $%
\left| \psi _{1}\right\rangle $ in terms of $\left| \psi ^{\prime
}\right\rangle $ and $\left| \psi \right\rangle ,$ we obtain
$I\otimes \Pi _{E_{1}}\left| \psi ^{\prime }\right\rangle =\alpha
\left( I\otimes \Pi _{E_{1}}\left| \psi \right\rangle \right) $
for some constant $\alpha$. It follows that $\left| \psi ^{\prime
}\right\rangle =\alpha \left( I\otimes
\Pi _{E_{1}}\left| \psi \right\rangle \right) +\beta \left| \chi
\right\rangle ,$ where $I\otimes \Pi _{E_{1}}\left| \chi \right\rangle =0$
and $\alpha ,\beta $ are constrained to ensure that $\left| \psi ^{\prime
}\right\rangle $ is normalized. Heuristically, Alice can pass Bob's test
with probability 1 whenever she submits a state $\left| \psi ^{\prime
}\right\rangle $ that is indistinguishable from $\left| \psi \right\rangle $
within the support of $E_{1}.$ Alice's probability of winning in this case
is $\left\langle \psi ^{\prime }\right| I\otimes E_{1}\left| \psi ^{\prime
}\right\rangle =\left| \alpha \right| ^{2}\left\langle \psi \right| I\otimes
E_{1}\left| \psi \right\rangle =\frac{1}{2}\left| \alpha \right| ^{2},$
which is maximized when $\beta =0$ and $\alpha =1/\sqrt{\left\langle \psi
\right| I\otimes \Pi _{E_{1}}\left| \psi \right\rangle }.$ This yields $P_{A}^{%
\text{thresh}}=1/2\left\langle \psi \right| I\otimes \Pi _{E_{1}}\left|
\psi \right\rangle =1/2\mathrm{Tr}\left( \rho \Pi
_{E_{1}}\right).$\hfill\nolinebreak\hfill$\blacksquare$

For proving properties 3 and 4, the following definition and
lemma are useful. (For simplicity we ignore degeneracy and
support issues which are easily incorporated but do not change any of our results.)

\noindent\emph{\textbf{Definition:}} Consider a vector $\left| \varphi \right\rangle \in
\mathcal{H}^{A}\otimes \mathcal{H}^{B},$ a linear operator $X$ on $\mathcal{H%
}^{A}$ and a linear operator $Y$ on $\mathcal{H}^{B}.$ $X$ and $Y$ are said
to be \emph{Schmidt equivalent under }$\left| \varphi \right\rangle $ if the
matrix elements of $X$ in the eigenbasis of $\mathrm{Tr}_{B}\left( \left|
\varphi \right\rangle \left\langle \varphi \right| \right) ,$ are the same
as the matrix elements of $Y$ in the eigenbasis of $\mathrm{Tr}_{A}\left(
\left| \varphi \right\rangle \left\langle \varphi \right| \right) .$\emph{\ }

\noindent \emph{\textbf{Lemma}}\cite{qic}: For a vector $\left| \varphi \right\rangle
\in \mathcal{H}^{A}\otimes \mathcal{H}^{B},$ and a positive operator $E$ on $%
\mathcal{H}^{B},$%
\[
\mathrm{Tr}_{B}\left( \left( I\otimes \sqrt{E}\right) \left| \varphi
\right\rangle \left\langle \varphi \right| \left( I\otimes \sqrt{E}\right)
\right) =\sqrt{\omega }D^{T}\sqrt{\omega },
\]
where $\omega \equiv \mathrm{Tr}_{B}\left( \left| \varphi \right\rangle
\left\langle \varphi \right| \right) $, $D$ is the operator on $\mathcal{H}%
^{A}$ that is Schmidt equivalent to $E$ under $\left| \varphi
\right\rangle , $ and $D^{T}$ is the transpose of $D$ with
respect to the eigenbasis of $\omega .$

\noindent \emph{\textbf{Proof of lemma:}}\textbf{\ }Suppose the bi-orthogonal
decomposition of $\left| \varphi \right\rangle $ is $\left| \varphi
\right\rangle =\sum_{j}\sqrt{\lambda _{j}}\left| e_{j}\right\rangle \otimes
\left| f_{j}\right\rangle .$ Taking the trace in terms of the basis $\left\{
\left| f_{i}\right\rangle \right\} ,$ we find
$\mathrm{LHS} = \sum_{j,k}\sqrt{\lambda _{j}\lambda _{k}}\left| e_{j}\right\rangle
\left\langle f_{k}\right| E\left| f_{j}\right\rangle \left\langle
e_{k}\right| .$ By definition, $\left\langle f_{k}\right| E\left|
f_{j}\right\rangle =\left\langle e_{k}\right| D\left|
e_{j}\right\rangle $ and $\left\langle e_{k}\right| D\left|
e_{j}\right\rangle =\left\langle e_{j}\right| D^{T} \left|
e_{k}\right\rangle .$ With some re-ordering of terms, we obtain
$\mathrm{LHS}=(\sum_{j}\sqrt{\lambda _{j}}\left|
e_{j}\right\rangle \left\langle e_{j}\right|
)D^{T}(\sum_{k}\sqrt{\lambda
_{k}}\left| e_{k}\right\rangle \left\langle e_{k}\right| ).$ Noting that $%
\sqrt{\lambda _{j}}$ and $\left| e_{j}\right\rangle $ are the eigenvalues
and eigenvectors of $\sqrt{\omega },$ we have the desired result.
\hfill $\blacksquare $

\noindent\emph{\textbf{Proof of Property 3:}}  Assume that Alice is honest.
Bob's most general cheating strategy can be implemented as follows. First,
he performs a measurement on system $B$ of a POVM $\left\{ E_{k}^{\prime
}\right\} ,$ which may have an arbitrary number of outcomes. With
probability $p_{k}^{\prime }=\left\langle \psi \right| I\otimes
E_{k}^{\prime }\left| \psi \right\rangle $ the outcome is $k$ and the state
of the total system is updated to $\left| \psi _{k}^{\prime }\right\rangle
=\left( I\otimes \sqrt{E_{k}^{\prime }}\left| \psi \right\rangle \right) /%
\sqrt{p_{k}^{\prime }}$. After the measurement, Bob can perform a unitary
transformation, $U_{k},$ on system $B$, the nature of which depends on the
outcome $k$ that was recorded. Finally, he must decide whether to announce $%
b=0$ or $1$ based on the result of the measurement, that is, he must decide
on a set $S_{0}$ of outcomes for which he will announce $b=0.$

Bob's probability of passing Alice's test given outcome $k$ is $\left|
\left\langle \psi _{0}\right| I\otimes U_{k}\left| \psi _{k}^{\prime
}\right\rangle \right| ^{2}$, so his probability of winning the
coin flip is $P_{B}=$ $\sum_{k\in S_{0}}p_{k}^{\prime }\left|
\left\langle \psi _{0}\right| I\otimes U_{k}\left| \psi
_{k}^{\prime }\right\rangle \right| ^{2}.$ We must maximize this
with respect to variations in $\left\{ E_{k}^{\prime }\right\}
,\left\{ U_{k}\right\} ,$ and $S_{0}.$ By Uhlmann's theorem
\cite{jozsa}, $\sup_{U_{k}}\left| \left\langle \psi _{0}\right|
I\otimes U_{k}\left| \psi _{k}^{\prime }\right\rangle \right|
^{2}=F\left( \sigma
_{0},\sigma _{k}^{\prime }\right) ^{2},$ where $\sigma _{b}\equiv \mathrm{Tr%
}_{B}\left( \left| \psi _{b}\right\rangle \left\langle \psi
_{b}\right|
\right) ,$  $\sigma _{k}^{\prime }\equiv \mathrm{Tr}_{B}\left( \left|
\psi _{k}^{\prime }\right\rangle \left\langle \psi _{k}^{\prime }\right|
\right) $ and $F(\omega ,\tau )\equiv \mathrm{Tr}|\sqrt{\omega
}\sqrt{\tau }|$ is the fidelity. Thus we need to
compute $P_{B}^{\max }=\sup_{\left\{ E_{k}^{\prime }\right\}
,S_{0}}\sum_{k\in S_{0}}F\left( \sigma _{0},p_{k}^{\prime }\sigma
_{k}^{\prime }\right) ^{2}$. Since the fidelity squared is always
positive, $\sum_{k\in S_{0}}F\left( \sigma _{0},p_{k}^{\prime
}\sigma _{k}^{\prime }\right) ^{2}\le \sum_{k}F\left(
\sigma _{0},p_{k}^{\prime }\sigma _{k}^{\prime }\right) ^{2}.$ This implies
that the optimal $S_{0}$ is the entire set of indices: no matter
what the outcome $k$ of Bob's measurement, he should announce bit
$0.$ Moreover, by the concavity of the fidelity squared
\cite{jozsa}, we have $\sum_{k}F\left( \sigma _{0},p_{k}^{\prime
}\sigma _{k}^{\prime }\right) ^{2}\le F(\sigma
_{0},\sum_{k}p_{k}^{\prime }\sigma _{k}^{\prime })^{2}=F\left(
\sigma _{0},\sigma \right) ^{2},$ where $\sigma \equiv
\mathrm{Tr}_{B}\left( \left| \psi \right\rangle
\left\langle \psi \right| \right) $. This upper bound is saturated if Bob
makes no measurement upon system $B.$ Using the definition of $\left| \psi
_{0}\right\rangle $ and the lemma, we find that $\sigma _{0}=2\sqrt{\sigma }%
D_{0}^{T}\sqrt{\sigma },$ where $D_{0}$ is Schmidt equivalent to $E_{0}$
under $\left| \psi \right\rangle .$ Thus, we can write $P_{B}^{\max }=$ $%
F\left( 2\sqrt{\sigma }D_{0}^{T}\sqrt{\sigma },\sigma \right) ^{2}=F\left( 2%
\sqrt{\sigma }D_{0}\sqrt{\sigma },\sigma \right) ^{2},$ where the second
equality follows from the fact that $X^{T}
$ and $X$ have the same eigenvalues. By the isomorphism between $%
\mathcal{H}^{A}$ and $\mathcal{H}^{B}$ induced by Schmidt equivalence under $%
\left| \psi \right\rangle $, we have
$P_{B}^{\max }=F\left( 2\sqrt{\rho }E_{0}\sqrt{\rho },\rho
\right) ^{2}$.  Finally, by the definition of the fidelity, we
have $P_{B}^{\max }=2(\mathrm{Tr}\sqrt{\rho E_0 \rho})
^{2}.$\hfill $\blacksquare $

\noindent\emph{\textbf{Proof of Property 4:}}  We seek to determine Bob's
maximum probability of winning assuming that his probability of
being caught cheating is strictly zero. The latter condition
constrains Bob's most general cheating strategy, described above,
to be such that he must always pass Alice's test whenever he
announces the outcome $b=0.$ That is, we require that
$\{E_{k}^{\prime} \}$, $\{ U_{k} \}$ and $S_{0}$ be such that
$I\otimes U_{k}\left|
\psi _{k}^{\prime }\right\rangle =\left| \psi _{0}\right\rangle $ for all $%
k\in S_{0}.$ The probability that Bob wins the coin flip is simply $%
\sum_{k\in S_{0}}p_{k}^{\prime },$ so we seek to determine $\sup_{\left\{
U_{k}\right\} ,\left\{ E_{k}^{\prime }\right\} ,S_{0}}\left( \sum_{k\in
S_{0}}p_{k}^{\prime }\right) ,$ where the optimization is subject to the
above constraint. We solve the optimization problem by establishing an upper
bound and demonstrating that it can be saturated.
We begin by using the definitions of $\left| \psi _{k}^{\prime
}\right\rangle $ and $\left| \psi _{0}\right\rangle $ to rewrite the
constraint equation as $\frac{1}{p_{k}^{\prime }}\left( I\otimes U_{k}\sqrt{%
E_{k}^{\prime }}\right) \left| \psi \right\rangle \left\langle \psi \right|
\left( I\otimes U_{k}\sqrt{E_{k}^{\prime }}\right) $ $=2\left( I\otimes
\sqrt{E_{0}}\right) \left| \psi \right\rangle \left\langle \psi \right|
\left( I\otimes \sqrt{E_{0}}\right) $ Tracing over $B$ and
applying the lemma provided above, we obtain $\sqrt{\sigma }(D_{k}^{\prime
})^{T}\sqrt{\sigma }=2p_{k}^{\prime }\sqrt{\sigma }D_{0}^{T}\sqrt{\sigma },$
where $D_{k}^{\prime }$ and $D_{0}$ are the Schmidt equivalent operators to $%
E_{k}^{\prime }$ and $E_{0}$ respectively. It follows that $\Pi _{\sigma
}D_{k}^{\prime }\Pi _{\sigma }=2p_{k}^{\prime }\Pi _{\sigma }D_{0}\Pi
_{\sigma },$which, by the isomorphism between $\mathcal{H}^{A}$ and $%
\mathcal{H}^{B}$ induced by Schmidt equivalence under $\left| \psi
\right\rangle $, implies $\Pi _{\rho }E_{k}^{\prime }\Pi _{\rho
}=2p_{k}^{\prime }\Pi _{\rho }E_{0}\Pi _{\rho }.$ Combining this with $%
\sum_{k\in S_{0}}E_{k}^{\prime }\le I,$ we obtain $\sum_{k\in
S_{0}}2p_{k}^{\prime }\Pi _{\rho }E_{0}\Pi _{\rho }\le \Pi _{\rho
},$ which in turn implies that $\sum_{k\in S_{0}}p_{k}^{\prime
}\le 1/2\lambda ^{\max }\left( \Pi _{\rho }E_{0}\Pi _{\rho
}\right) =1/2\lambda ^{\max }\left( \Pi _{\rho }E_{0}\right) .$
The upper bound can be saturated while satisfying the constraint
if Bob measures the POVM, $\left\{ E_{0}^{\prime
},E_{1}^{\prime }\right\} ,$ defined by $E_{0}^{\prime }=\Pi
_{\rho }E_{0}\Pi _{\rho }/\lambda ^{\max }\left( \Pi _{\rho
}E_{0}\right) ,$ and announces $b=0$ when he obtains the outcome
associated with $E_{0}^{\prime }$ \cite{fnote2}.
 Thus, Bob's threshold is $P_{B}^{\text{thresh}}=1/2\lambda ^{\max
}\left( \Pi _{\rho }E_{0}\right).$\hfill$\blacksquare$

The ordering of the authors on this paper was chosen by a coin
flip implemented by a trusted third party. TR lost.

\begin{acknowledgments}
RWS wishes to thank the group of Anton Zeilinger at the
University of Vienna for their hospitality during his visit.
This research was
supported in part by the Natural Sciences and Engineering
Research Council of Canada, the Austrian Science Foundation FWF,
the TMR programs of the European Union, Project No.
ERBFMRXCT960087, and by the NSA \& ARO under contract No.
DAAG55-98-C-0040.
\end{acknowledgments}


\begin{thebibliography}{9}
\bibitem{2wayCS} This is unfortunate since we believe that such a protocol could be used to
build WCF that is arbitrarily bias-resistant against both parties.
\bibitem{fnote2} An alternative proof of property 4 can be found by using the following result:
given two density operators $\omega$ and $\tau$, satisfying
$\Pi_{\omega}\ge \Pi_{\tau}$, the largest probability with which
$\tau$ can appear in a convex decomposition of $\omega$ is
$1/\lambda^{\max}(\omega^{-1}\tau)$. This result was first
pointed out to us by Michael Nielsen.
\end{thebibliography}
\end{document}